\documentclass[twocolumn]{aastex63}

\usepackage[utf8]{inputenc}
\usepackage{natbib}
\usepackage{graphicx}
\usepackage{amsmath}
\usepackage{booktabs}
\usepackage{longtable}
\usepackage{xspace}
\usepackage{hyperref}
\usepackage[T1]{fontenc}
\usepackage{lipsum}
\usepackage{comment}
\usepackage{xcolor}
\usepackage{enumitem}
\usepackage{multirow}
\usepackage{IEEEtrantools}
\usepackage{rotating}
\usepackage{wrapfig}
\usepackage{xspace}
\usepackage{makecell}

\newcommand{\teff}{\ensuremath{T_{\mathrm{eff}}}\xspace}
\newcommand{\logg}{\ensuremath{\log g}\xspace}
\newcommand{\feh}{[Fe/H]\xspace}
\newcommand{\vsin}{\ensuremath{v\sin i}\xspace}

\newcommand{\vmac}{\ensuremath{\zeta}\xspace}

\newcommand{\numax}{\ensuremath{\nu_{*,max}}\xspace}
\newcommand{\numaxxx}{\ensuremath{\nu_{\rm max}}\xspace}

\newcommand{\slope}{-2.30\xspace}
\newcommand{\slopeerr}{0.12\xspace}
\newcommand{\intercept}{5.32\xspace}
\newcommand{\intercepterr}{0.10\xspace}
\newcommand{\kms}{km s$^{-1}$}
\newcommand{\ms}{m s$^{-1}$}

\begin{document}

\title{Predicting Macroturbulence in F/G/K Dwarfs to 100 m/s Precision}

\author[0000-0001-8342-7736]{Jack Lubin}
\affiliation{Department of Physics \& Astronomy, University of California Los Angeles, Los Angeles, CA 90095, USA}

\author[0000-0003-0967-2893]{Erik A. Petigura}
\affiliation{Department of Physics \& Astronomy, University of California Los Angeles, Los Angeles, CA 90095, USA}

\author[0000-0003-1298-9699]{Kento Masuda}
\affiliation{Department of Earth and Space Science, Osaka University, Osaka 560-0043, Japan}

\author[0000-0003-1312-9391]{Samuel Halverson}
\affiliation{Jet Propulsion Laboratory, California Institute of Technology, 4800 Oak Grove Drive, Pasadena, CA 91109, USA}


\author[0000-0002-9751-2664]{Isabel Angelo}
\affiliation{SETI Institute, 339 Bernardo Ave, Suite 200, Mountain View, CA 94043, USA}

\author[0000-0001-8832-4488]{Daniel Huber}
\affiliation{Institute for Astronomy, University of Hawai`i, 2680 Woodlawn Drive, Honolulu, HI 96822, USA}

\author[0000-0002-4677-8796]{Michael L. Palumbo III}
\affiliation{Center for Computational Astrophysics, Flatiron Institute, 162 Fifth Avenue, New York, NY, USA}

\author[0000-0002-7846-6981]{Songhu Wang}
\affiliation{Department of Astronomy, Indiana University, 727 East 3rd Street, Bloomington, IN 47405-7105, USA}

\author[0000-0002-0376-6365]{Xian-Yu Wang}
 \altaffiliation{Sullivan Prize Postdoctoral Fellow}
\affiliation{Department of Astronomy, Indiana University, 727 East 3rd Street, Bloomington, IN 47405-7105, USA}

\begin{abstract}
We leverage the high resolution and spectral stability of the Keck Planet Finder (KPF) spectrograph to investigate macroturbulence broadening via the stellar cross-correlation function (CCF). As our calibration sample, we use main sequence benchmark slow-rotators (\vsin $<$ 4 {\kms}) where rotation rates were derived from the most robust asteroseismic mode splitting measurements, independent from spectral line broadening. We fit a linear relationship for macroturbulence as a function of derived \numaxxx, the frequency of maximum power due to stellar oscillations, with an RMS scatter of 90 {\ms}. Previous studies have calibrated macroturbulence against Teff, but we find \numaxxx to be a lower dispersion predictor by a factor of 4 indicating a deeper relationship between \numaxxx and convective motions.

\end{abstract}

\section{Introduction}
\label{Intro}

The rotation rate of a star is a fundamental parameter as well as a tracer for many facets of stellar and exoplanetary astrophysics. Rotation is related to the stellar dynamo, magnetic fields, and stellar activity \citep{Noyes1984}. Rotation rates as a function of time probe stellar evolution and are the basis for the field of gyrochronology \citep{Bouma2023}. Furthermore, the relationship between the stellar rotation axis and the planet's orbital inclination, i.e. the ``stellar obliquity angle'' is a key observable feature of a planetary system's architecture and therefore its formation history \citep{Winn2005}.

Due to the differential velocity from the approaching and receding limbs of a star, the projected rotational velocity of the star by the stellar inclination angle, \vsin, is imprinted into the spectrum through the broadening of lines. However, rotation is not the only source of broadening in the spectral lines. Thermal broadening occurs from the projected Gaussian distribution of velocities of individual atoms in the gas, and pressure broadening occurs from the distribution of interaction separation between absorber and perturber in collisions between atoms \citep{Gray2005}. Additionally, microturbulence broadens lines due to the motions of the material in convection where the size of the material and scale of motion is smaller than the optical depth. And macroturbulence, $\zeta$, relating to the wholesale movement of groups of granules in the photosphere due to convection on scales larger than optical depth, also broadens spectral lines. 

Even with high resolution spectrographs (R$\sim$100k), when broadening from \vsin is smaller than that from macroturbulence, microturbulence, and/or the instrument's line spread function (LSF) its measurement is limited by one’s ability to accurately model these other terms \citep{Masuda2022}. To highlight the challenges, in the literature very slow rotators ($<$2 {\kms}, like our Sun at 1.9 {\kms} \citealp{Gray2005}) typically receive only an upper limit on \vsin. With new, stabilized spectrographs, the LSF can be well characterized to the few {\ms} level. In the error budget, broadening from $\zeta$ is often greater than that of natural, thermal, and pressure broadening and so with deliberate choosing of the line list, the latter sources are often ignored. However, $\zeta$ itself is often not well characterized, becoming the primary impediment to a precise \vsin measurement in the slow rotation regime. Beyond this, macroturbulence both degrades the precision of radial velocity (RV) measurements and it is an unknown parameter in Rossiter-McLaughlin (RM) measurements \citep{Rossiter1924,McLaughlin1924}. A robust understanding of macroturbulence will open the door to more advances in these sub-fields as well.

Prior works in the literature have developed $\zeta$ relationships with effective temperature (\teff) \citep{VF2005, Brewer2016}, but rely on the assumption of \vsin = 0 {\kms} for the stars in the sample. In \citet{Doyle2014} (hereafter D14), $\zeta$ was measured simultaneously with \vsin on a selection of stars with \vsin priors from asteroseismology. In this work, we measure $\zeta$ on a similar set of stars to D14; however we make use of higher resolution spectra (R$\sim$100k) and operate in the cross-correlation (CCF) space. 

The paper is organized as follows. In \S\ref{Sample} we introduce the sample of stars with which we demonstrate our new technique, as well as the more stringent sub-sample on which we derive our macroturbulence relationship. We also describe the observations of these stars. Next, \S\ref{Methods} describes the construction of a KPF synthetic spectra generator which we use to measure macroturbulence in our primary experiment, and in \S\ref{Results} we report the results and derive a linear relationship between macroturbulence and \numaxxx, the frequency of maximum oscillation power in the photometric power spectrum. In \S\ref{Discussion} we contextualize our results, motivate the use of \numaxxx over \teff to set our relationship, and discuss future directions of this work. Finally in \S\ref{Conclusion} we conclude.

\begin{figure}[t!]
\centering\includegraphics[width=0.5\textwidth]{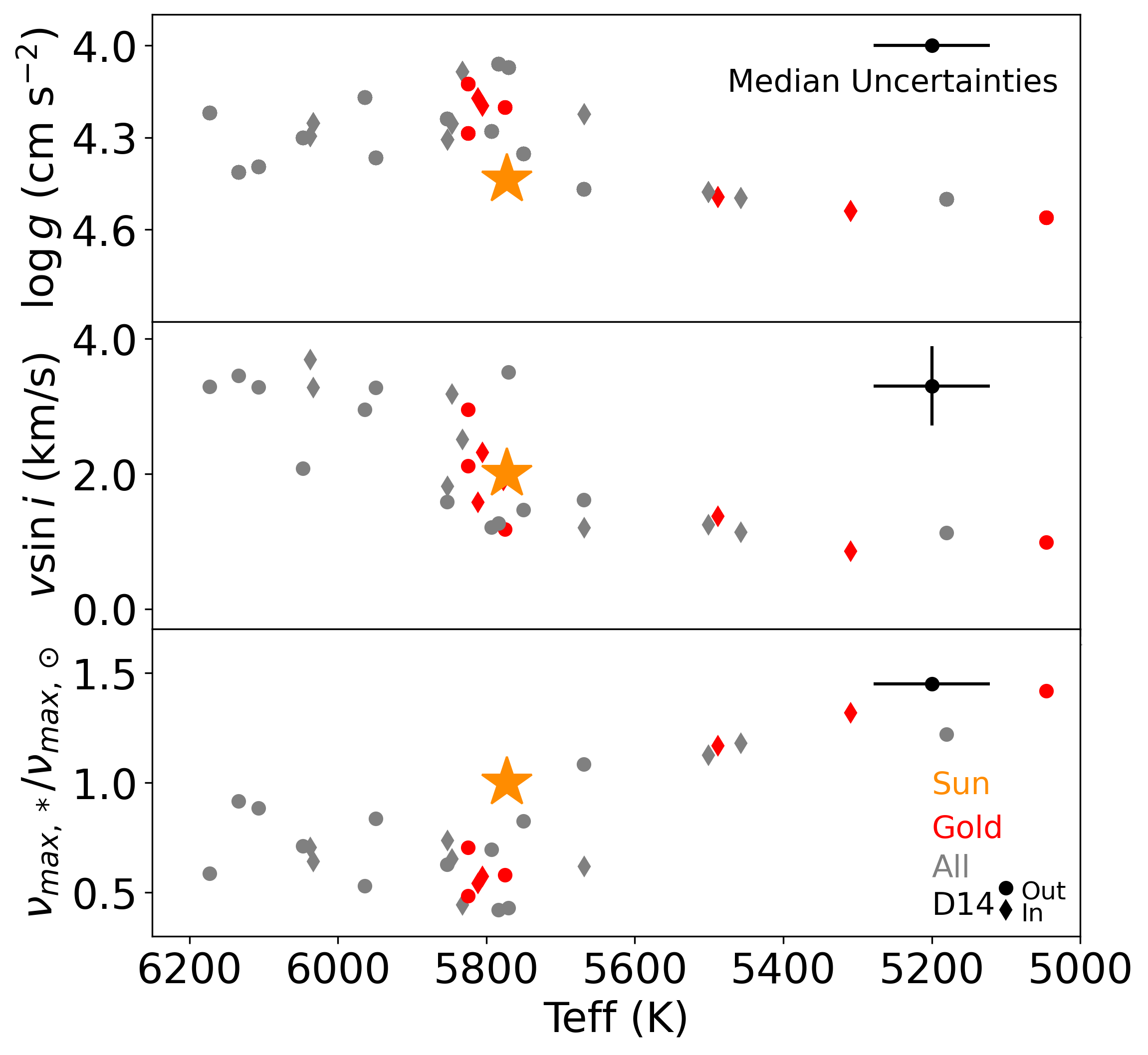}
\caption{Our stellar sample displayed as a function \logg (top), \vsin (middle), and \numax (bottom) against \teff. Representative error bars on each parameter are shown within the panel, including in the y-dimension. Stars in our sample that overlap with the D14 sample are shown as diamonds. Stars in the ``gold'' sample are shown in red. The orange star indicates our Sun, which is included in the gold sample.} 
\label{hrdiagram}
\end{figure}

\section{Sample}
\label{Sample}

We compiled our sample from a pool of stars which each have precisely measured \vsin from a spectrum-independent method. In particular, we used the \citet{Hall2021} pool of bright Kepler stars, which includes precise masses, surface gravities (\logg), and the frequency width of the mode splitting by the stellar inclination angle ($\nu_s \sin i$) as measured from asteroseismology. We computed \vsin via

\begin{equation}
    \vsin =2\pi \sqrt{\frac{G M}{g}} \nu_s\sin i
\end{equation}

\noindent and propagated uncertainties. We cut the pool to keep only stars with \vsin $<4$ {\kms}. There are 46 stars that passed this cut, of which we ultimately observed 30. We note that the seismic oscillation amplitudes scale with luminosity and mass, and therefore the measurement is biased against low mass stars and un-evolved main sequence stars which is reflected in our sample, see Figure \ref{hrdiagram}. The stellar parameters adopted from \citet{Hall2021} are reproduced in Table \ref{results_table}. The median uncertainties in our sample are $77$ K in \teff, $0.006$ dex in \logg, and $0.095$ dex in metallicity. 

We used \teff and \logg to compute \numaxxx via the scaling relationship \citep{Brown1991, Kjeldsen1995}: 

\begin{equation}
    \frac{\nu_{*_{max}}}{\nu_{\odot_{max}}} = \Big(\frac{g_*}{g_{\odot}}\Big)  \Big(\frac{\mathrm{\teff}_*}{\mathrm{\teff}_{\odot}}\Big)^{-1/2}
\label{numaxEQ}
\end{equation}

where solar values are 5772 K in \teff and 4.44 dex in \logg. We note this relationship itself depends on the assumed linear relationship between \numaxxx and the acoustic cutoff frequency ($\nu_{c}$) first pointed out in \citet{Brown1991}, and later physically connected in \citet{Belkacem2011}.

\begin{figure*}[t!]
\centering\includegraphics[width=\textwidth]{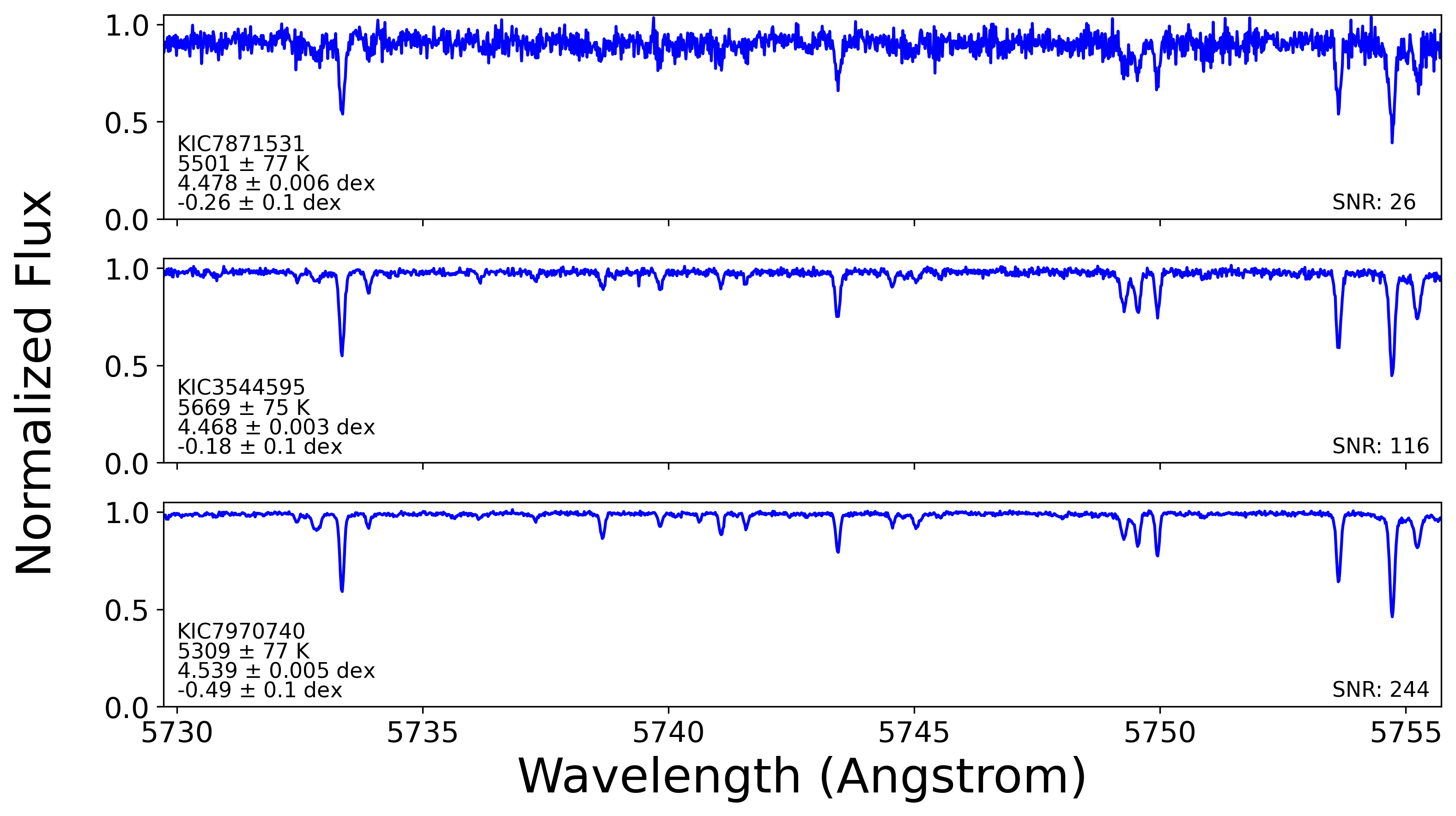}
\caption{Examples of spectra from our observations with KPF. We show the lowest (top), median (middle) and highest (bottom) SNR observations.} 
\label{sample_spectra}
\end{figure*}

We observed these stars with the Keck Planet Finder (KPF) \citep{Gibson2024}. KPF is a fiber-fed, high resolution R$\sim$98,000 spectrograph. It has exquisite spectral stability which essentially fixes the instrument profile at the scales of our measurements. Over the 4 months of data collection between February and July of 2025, the FWHM of the LSF was stable to 20 {\ms} as determined through investigating calibration images of the laser frequency comb. This is an important factor for all subsequent analysis. We executed observations of targets through the KPF-Community Cadence (KPF-CC) queue \citep{Handley2024, Lubin2025}. Excluding the Sun (SNR = 586), the SNR of our observations ranges from 26 to 244, with a median of 117, see Figure \ref{sample_spectra}.  As we will show, we are able to obtain good results even on lower SNR spectra because, through our CCF technique, we are not photon limited but rather limited by the shapes of the lines themselves.

We processed the spectra through the standard KPF Data Reduction Pipeline (DRP) \footnote{\url{https://github.com/Keck-DataReductionPipelines/KPF-Pipeline}} and we made use of the L1 data product, which is a one-dimensional, wavelength calibrated spectrum. In reviewing the spectra and associated data products, we discerned that KIC9025370 is a SB2 binary star through its clear secondary minimum in the CCF. While the companion is separated widely in velocity space ($\Delta$v $\simeq$35 {\kms}), we excluded it to keep our sample to only single stars. We further added the Sun to our sample, selecting a spectrum out of a vetted time-series from \citet{Rubenzahl2023}. This left a final sample of 30 stars, of which 12, including the Sun, overlap with the sample from D14. This split will allow us to ground our own method's determination of macroturbulence with accepted literature while also applying it to new targets. 

Of these stars, not all asteroseismic measurements are of equal quality. In \citet{Kamiaka2018}, the authors describe data quality cuts on the Kepler asteroseismic sample. Through simulations, they define reliable asteroseismic mode-splitting measurements, and the resulting stellar inclination angles and therefore \vsin values, as stars whose data pass two cuts. First, the ``Height-to-Background Ratio'' (HBR) value represents a SNR-like quantity to the mode-splitting detection and they recommend trusting measurements where HBR $>$ 3. Second, they define a dimensionless factor relating to the stellar rotation and intrinsic width of the Lorentzian-like mode peaks, $d\nu_* / \Gamma$, which they state should exceed a value of 0.5 to be considered a reliable measurement. Applying these two cuts to the 29 Kepler stars in our sample leaves 8 stars (to which we add the Sun). We denote this 9 star sub-sample as our ``gold'' sample, which we will use to derive our macroturbulence relationship.

\section{Demonstration of Method}
\label{Methods}

\begin{figure}[t!]
\centering\includegraphics[width=0.5\textwidth]{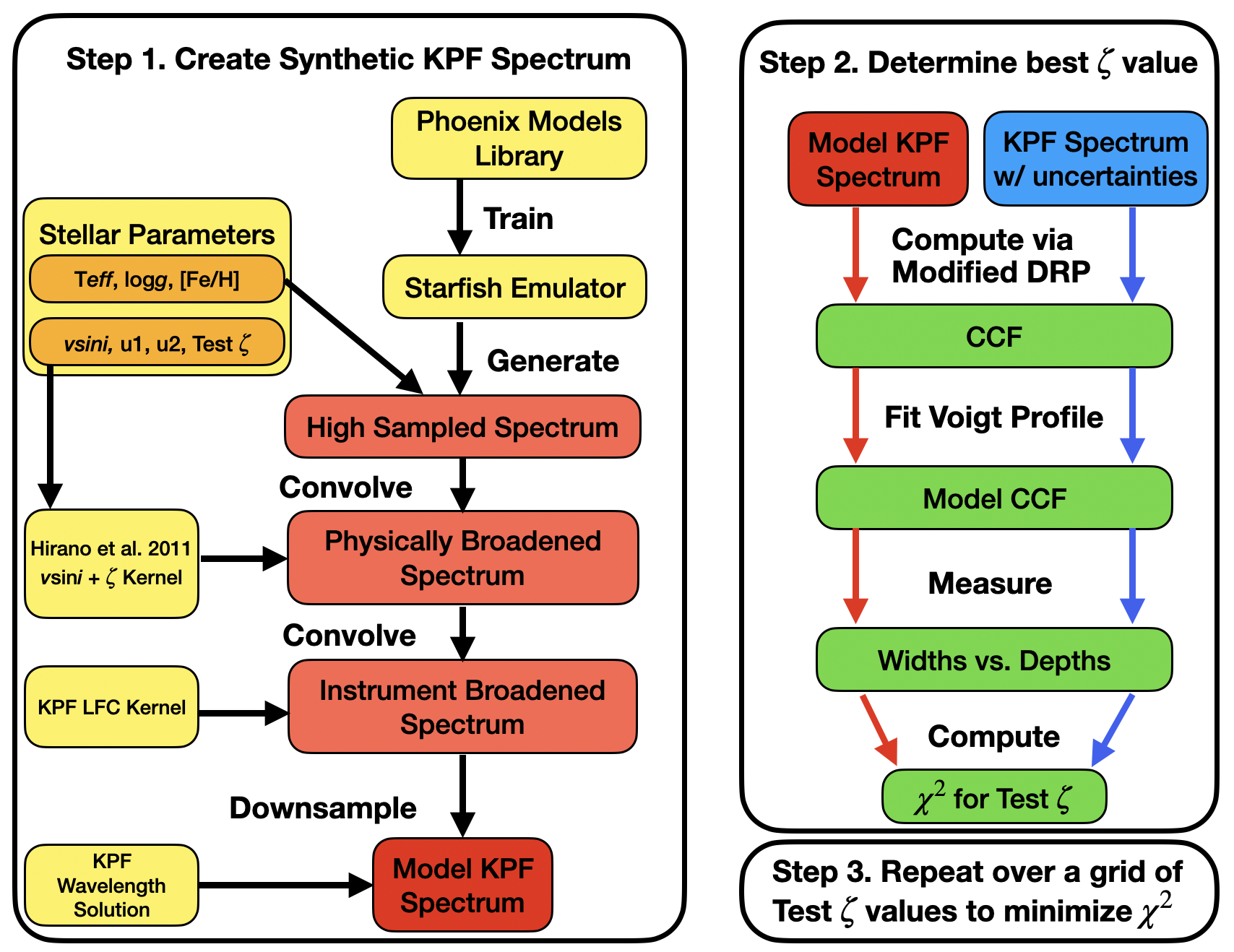}
\caption{An illustrative schematic of our pipeline to produce synthetic KPF spectra and compute macroturbulence values.} 
\label{schematic}
\end{figure}

In order to measure the macroturbulence of a star, we constructed a pipeline which generates synthetic KPF spectra from which we can compute its CCF. See Figure \ref{schematic} for a schematic diagram of the pipeline's overview. 

\subsection{Pipeline}
\label{Pipeline}

We started by creating an emulator using \texttt{Starfish} \citep{Czekala2015} which interpolates between spectra in the Phoenix grids \citep{Husser2013} to generate a spectrum with a user defined \teff, \logg, and \feh . We note that the Phoenix spectra contain microturbulence broadening. Our emulator was designed to produce spectra in the KPF Green bandpass, 445-600 nm, within parameter ranges of 3500-7000 K in \teff, 4.0-5.0 dex in \logg, and -0.5 to +0.5 dex in metallicity. We modified \texttt{Starfish} so that our emulator does not perform any instrumental broadening. Instead, we will later use an empirical KPF instrument profile. We produced spectra one KPF order at a time, choosing to work with orders 28 through 32 of the KPF Green Chip ($\sim$560-588~nm). The results from these orders matched those from D14 best, we believe due to the lower rate of blends, which can effectively act as wider lines and imprint as higher macroturbulence values.

The emulator produced spectra at average $\Delta$v $\simeq$ 0.020 {\kms}, which we resampled using Starfish's built-in functionality to be uniform in log lambda. We chose $\Delta$v to be 0.25 {\kms}, matching that of the KPF DRP step size when computing the CCF, as described below. This still over-samples KPF's native resolution (not uniform in log lambda, but $\Delta$v $\sim$ 0.9 {\kms}). 

\begin{figure*}[t!]
\centering\includegraphics[width=1\textwidth]{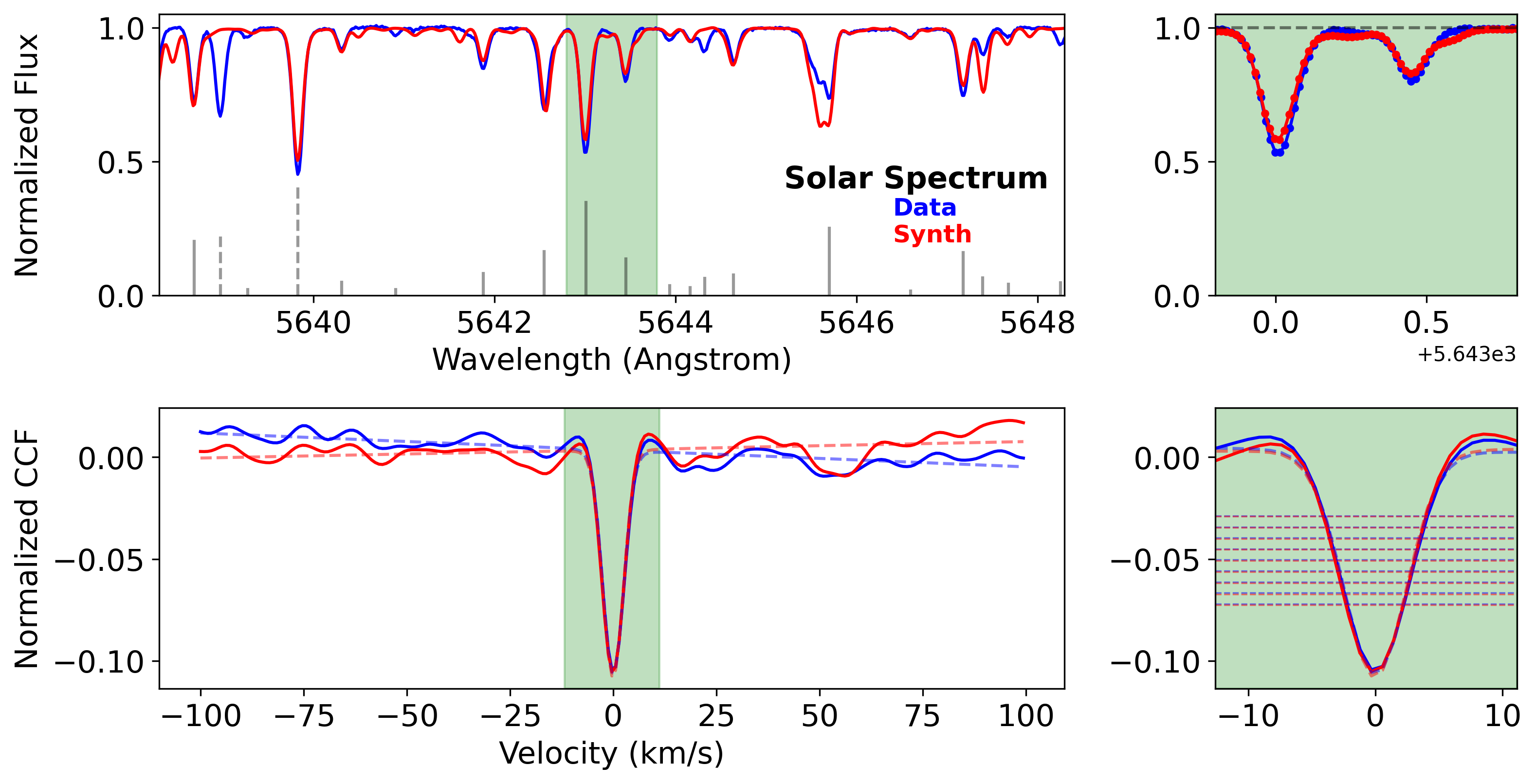}
\caption{\textbf{Top:} Our KPF spectrum of the Sun and its CCF (blue) compared to our Phoenix/Starfish interpolator synthetic spectrum generator and its CCF (red). The synthetic spectrum was produced with median values from Table \ref{results_table}, including our measured macroturbulence. The vertical gray lines indicate the locations and weights of the delta functions that make up the line list mask for the CCF calculation (dashed lines exist in the ESPRESSO G9 mask but were cut from our custom line list). \textbf{Bottom:} The orders weighted-sum, normalized CCFs of the two spectra. The right panels show zoom ins from the left panels within the green shaded regions. The bottom right figure contains horizontal dashed lines corresponding to the depths in the CCF that were tested to compare widths as a function of depth.} 
\label{synth_spectrum}
\end{figure*}

We then applied the following broadening mechanisms. First, we accounted for both rotation and the radial-tangential macroturbulence model (which must be treated simultaneously, see \citet{Gray2005} Chapter 17) using the \cite{Hirano2011} kernel. This model requires quadratic limb-darkening parameters, which we fit for on a star-by-star basis using \texttt{LDTk} \citep{Parviainen2015} and a boxcar bandpass equal to that spanning the 5 KPF orders used in our experiment. 

Second, we applied broadening from the instrument profile. KPF's LSF stability essentially fixes the instrument profile at the $\sim$20 {\ms}-level, allowing for confidence that our observations taken over the course of months all have essentially the same instrumental broadening. KPF's octagonal fiber is sliced into three traces, with the second trace cutting the middle of the octagon into a rectangle, and the outer (first and third) traces are symmetric trapezoids. We chose to only work with the rectangular second science trace as this makes an instrument profile that is a Gaussian convolved with a Top-hat. Note that the SNRs quoted in this work were computed using only the data in this middle trace. To match the as-built instrumental profile, we used the CCF of a calibration spectrum of the KPF Laser Frequency Comb (LFC). This CCF is computed with a finite width mask, with the same width as used for stellar RVs. We subtracted off a small bias from the detector, then normalized it before convolving it with the synthetic (and already astrophysically broadened) spectrum. 

Third, we continuum normalized the spectrum by a simple procedure. We divided the full synthetic spectrum into 10 equal size bins, approximately 150 \AA\, wide. We took the maximum flux value within each bin and paired it with the median wavelength in the bin to create a crude function that traces the upper envelope of the spectrum. We then interpolated this function at every wavelength before dividing it from the synthetic fluxes. The continuum normalizing process has no impact on the shapes/widths of the lines. It is solely performed for visual ease of comparison between synthetic spectra and their real counterparts. 

Fourth, within each KPF order of interest, we down-sampled the spectrum onto the KPF wavelength solution. We preserved flux by integrating under all the discrete wavelengths of the synthetic spectrum that fall closest to a given KPF wavelength value. This completed the process of producing a synthetic spectrum with realistic astrophysical and instrumental broadening terms. 

With this spectrum, we computed the CCF using a modified version of the KPF DRP. We computed the CCF on a velocity grid spaced to make step sizes equal to the velocity throw from one pixel to the next. This more easily allowed for the propagation of flux errors to CCF uncertainties. We used the same ESPRESSO G9 line list mask on all stars, electing to remove certain lines from the mask. Based on our Solar spectrum, we removed all lines which are deeper than 0.5 normalized flux in an effort to avoid saturated lines. Again using the solar spectrum, and a synthetic solar spectrum model from our pipeline, using \vsin = 1.9 km/s and \vmac = 3.1 km/s \citep{Gray2005} for astrophysical broadening, we additionally removed all lines where the percent difference between the depth of the real line and the synthetic line was greater than 20\%, giving the synthetic CCFs more consistency with the real CCFs. After cutting out the bad lines, our mask is left with 397 of the 500 original lines (80\%) in the orders of interest. Lastly, we normalized the resulting CCFs.

In Figure \ref{synth_spectrum}, we show an example comparing the KPF spectrum of the Sun to a synthetic spectrum from our pipeline. The synthetic spectrum was produced using literature median values for \teff, \logg, \feh, \vsin, as well as using $\zeta$ from our best fit value (see below). We similarly show the resulting CCFs for both real and synthetic spectra. Instead of using the DRP's Level 2 CCFs, we ran the real spectrum's wavelengths and fluxes through our modified CCF code. When computing these real CCFs, we used the raw DRP Level 1 fluxes which have not been de-blazed or continuum normalized. 

We note good qualitative agreement between the real and synthetic CCFs while also acknowledging quantitative disagreements. Generally, the spectra are similar, although there are many wavelengths where the synthetic fluxes deviate from the real fluxes by far more than the real error bars. This comes out of two noticeable features of the synthetic spectra: there are lines in the real spectrum that are completely missing in the synthetic one, and vice-versa, there are lines in the wrong location, and often line depths do not well match those of the real KPF spectrum. These differences are inherent to the Phoenix models themselves. All assumptions and limitations of the Phoenix models will propagate to our work here. We reiterate that we have minimized the inconsistencies by careful line list selection and so the CCFs of our synthetic spectra to real spectra are in good agreement. The CCFs are not sensitive to any one line in particular, but rather the full the set of lines in the mask. 

\subsection{CCF Width as a Function of CCF Depth}
\label{width_v_depth}

The Full Width at Half Max (FWHM) of the CCF is a well-established metric. Here we extend it by measuring width of the CCF at many depths in an effort to characterize the shape of the broadening from macroturbulence. 

To prepare for these measurements, we first performed a linear interpolation of the CCF onto a fine grid of velocities so that at each depth of the CCF there would be a CCF value symmetric across the 0 {\kms} vertical line. Next we standardize our definition of the depth within the CCF. The CCF represents the average line shape, and all lines are formed out of the continuum of the spectrum. It is this continuum, which has an analogous form in the CCF, that should define the top of the CCF. We chose to fit a Voigt model to the CCF, which includes parameters for the line amplitude, its center in velocity space, the Gaussian broadening $\sigma$, the Lorentzian broadening $\gamma$, and the continuum level. This is intended to be a simple parametric model of the CCF, not a physically realistic one. We additionally added a parameter to fit for the slope in the continuum. When the CCF mask has more lines on either the blue or red side of the peak of the blaze function, the continuum of the CCF is tilted. This has a negligible affect on our CCF width measurements, but if untreated it can affect the depths that are assigned to each width measurement. Using the fit's continuum and minimum depth levels, we have standardized the top and bottom of the CCF, respectively.

We then sliced the CCF horizontally at depths between 30\% and 70\% of the total depth, in increments of 5\% to create a grid of depth values with associated widths. We did not start at 0\% depth, or the very bottom of the line, as the width there is zero. In this work we are most sensitive to macroturbulence at the edges of the line. The continuum of the spectrum is less smooth than the lines and the CCFs reflect this. Therefore, we elected to not measure the widths at the top of the CCFs either. For the KPF data, which has flux uncertainties that are propagated to CCF uncertainties, we further propagated to uncertainties on the measured widths through a bootstrap resampling. For each CCF value and uncertainty, we drew a new CCF value from a Gaussian with a mean at the original CCF value and with sigma equal to the uncertainty. We then refit the Voigt profile and measured the widths again. We repeated this 100 times and used the mean and standard deviation of the 100 widths at each depth to serve as the measured widths and uncertainties for the spectrum. 

\subsection{Derivation of Macroturbulence}
\label{derivation}

To derive a macroturbulence relationship using the CCFs, we performed the following test. For each star in our sample, we produced a synthetic spectrum with the median literature values for \teff, \logg, \feh, and \vsin as well as an assumed value for $\zeta$. We repeated this on an order by order basis, computing the CCF each time. With each individual order's CCF, we computed a weighted sum CCF using the order weights from the KPF DRP, normalizing the weights before applying them to the CCFs. Finally, we computed width as a function of depth for the normalized, weighted-sum CCF of this synthetic spectrum. We then computed the widths, and their uncertainties, as function of depths for the real KPF spectrum, having similarly first computed the weighted-sum and normalized CCF for the matching orders.

As the synthetic spectrum is errorless, so too are the resulting CCF and computed widths. We computed the reduced $\chi^2$ between the synthetic CCF widths and the real CCF's uncertain widths. This goodness of fit is assigned to the macroturbulence value that was used to produce the spectrum. We continued this process, generating many synthetic spectra for the same star, each with the same \teff, \logg, \feh, \vsin, over a grid in $\zeta$ space from 0 to 6 {\kms} in steps of 0.1 {\kms}. The  $\zeta$ value corresponding to the minimum of the reduced $\chi^2$ curve is assigned as the best-fit macroturbulence value for the star. 

\begin{figure}[t!]
\centering
\includegraphics[width=0.5\textwidth]{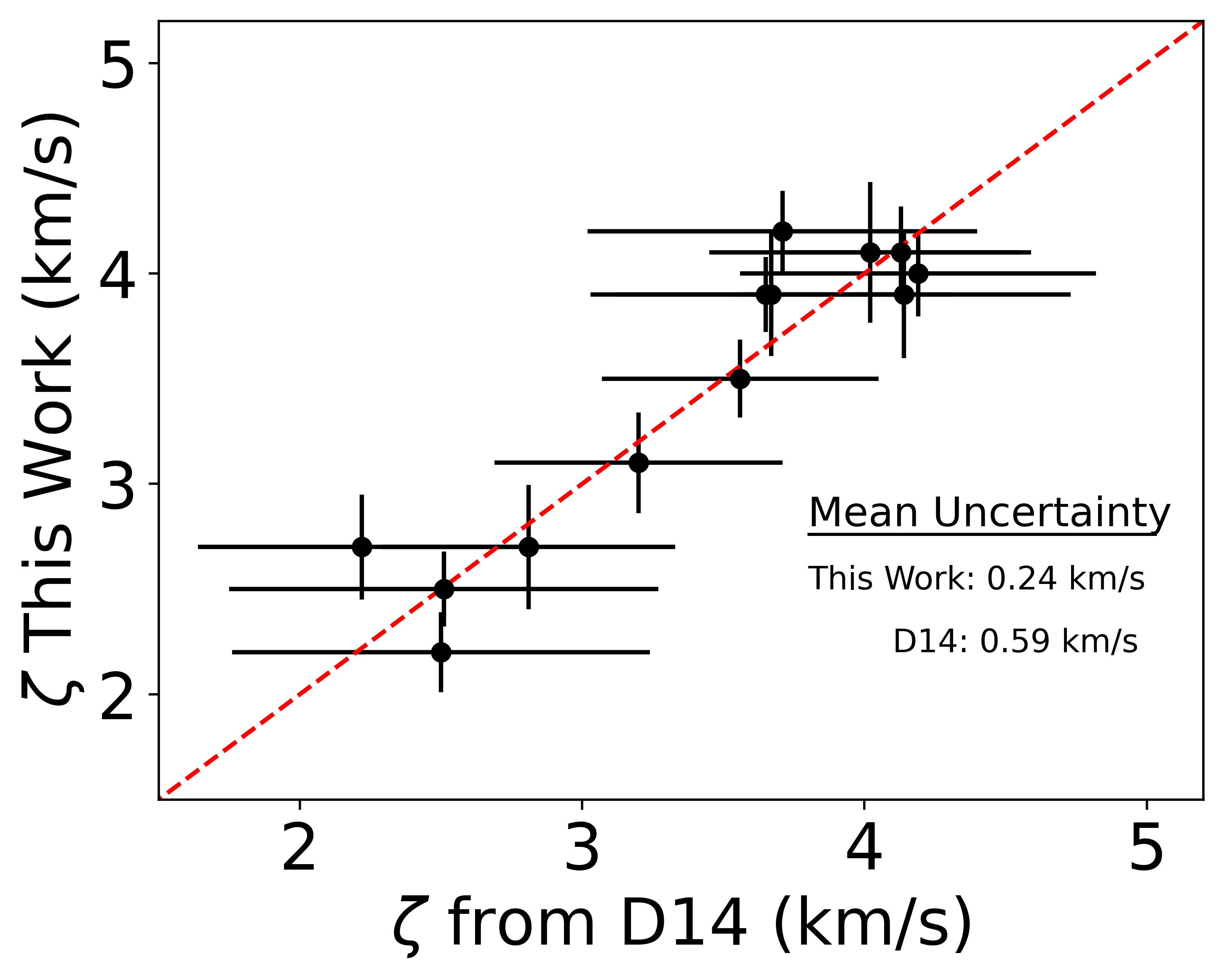}
\caption{A direct comparison of the macroturbulence values from our method with that of the method from D14 in the overlapping sub-sample.}
\label{one2one}
\end{figure}

In order to characterize the uncertainty in this best-fit macroturbulence value, we performed a bootstrap resampling of the data. For a given star, we repeated the process above in its entirety but instead of using all 5 orders once, we randomly drew from the orders 5 times with replacement, re-normalizing weights appropriately. We simultaneously drew a new \vsin value out of the literature posterior in an effort to encode its uncertainty into our macroturbulence uncertainty, as generally an increase in the \vsin value causes a decrease in the resulting macroturbulence value and vice-versa. We repeated the random draw 100 times per star. The standard deviation in the macroturbulence values for a given star was assigned to the best-fit macroturbulence's uncertainty for that star. 

\begin{figure*}[t!]
\centering
\includegraphics[width=\textwidth]{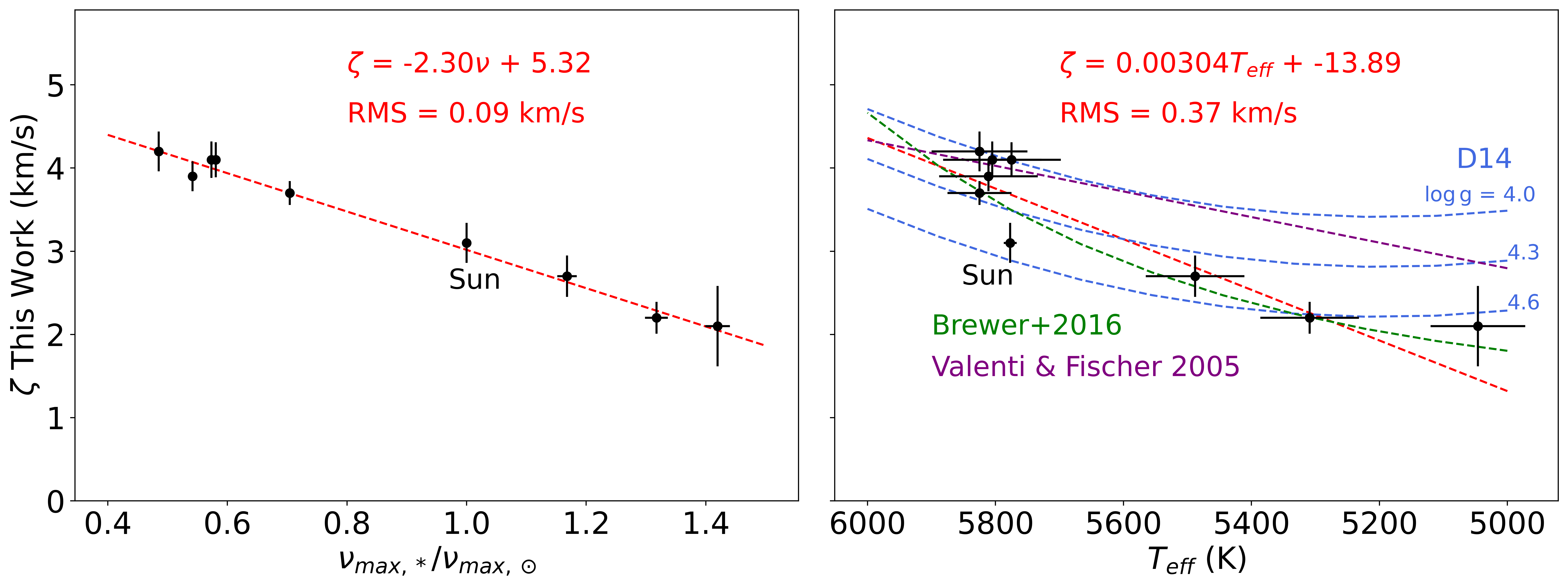}
\caption{Macroturbulence as a function of \numaxxx (left) and as a function of \teff (right). Only the ``gold'' sub-sample shown. The red dashed lines are our best fits from Orthogonal Distance Regression, accounting for uncertainties in both dimensions. Uncertainties on macroturbulence come from bootstrapping. Our fit against \numaxxx has a $\chi^2_{reduced}$ of 0.26, while our fit against \teff alone has a $\chi^2_{reduced}$ of 1.72. We show also the D14 model overlaid, which found an RMS of 0.37, as well as other relationships that are dependent only on \teff. We find that \numaxxx is the strongest predictor of macroturbulence.}
\label{relationship}
\end{figure*}

These results hold even under small changes in the assumed stellar parameters of the stars in the sample. We reran the entire pipeline three times under different scenarios: adding 100 K to the nominal \teff, adding 0.1 dex to the nominal \logg, and adding 0.1 dex to the nominal \feh. In comparing the macroturbulence values in each these tests to the nominal values, we find agreement within the 1$\sigma$ macroturbulence uncertainties in 91\% of tests. With the average error on \teff being 77 K, average error on \logg being 0.006 dex, and average error on \feh being 0.1 dex, our results are robust to typical uncertainties in stellar parameters. Additionally, incorrect limb darkening parameters have negligible effect on the resulting macroturbulence values. We investigated the effect of biases in the quadratic limb darkening coefficients, u1 and u2, by changing $u_1$ by 0.2 and $u_2$ by 0.1 (approximately $33\%$ changes) from each star's \texttt{LDTk} computed values and found no change in the resulting best fit macroturbulence value.

\section{Macroturbulence Relationship}
\label{Results}

In comparing our $\zeta$ values to those from D14 in the sub-sample of our stars that overlap, we find our results to be consistent. The RMS scatter about the one to one line is 0.25 {\kms}, see Figure \ref{one2one}. We note that our solar macroturbulence also matches literature values well. We find $\zeta$ = 3.1 $\pm$ 0.24 {\kms} where \citet{Gray1977} found of 3.1 $\pm$ 0.1 {\kms} and D14 found 3.2 {\kms} with a range of uncertainties depending on the resolution of the instrument used. While consistent with past literature, we also highlight that our method's uncertainties on individual $\zeta$ values are smaller by more than a factor of two. This is most likely because D14 used only 20 spectral lines for their analysis while we are using 397 via the CCF computation. 

Our main results are shown in Figure \ref{relationship}, where we plot $\zeta$ against \numaxxx and against \teff. Values are recorded in Table \ref{results_table}. We fit an Orthogonal Distance Regression (ODR) through our data, so as to take into account uncertainties in both the x and y dimensions, in order to measure the slope and intercept of a line through our results. We find our macroturbulence model to be the line that best fits our gold sample data via \numaxxx relationship as

\begin{equation}
\begin{aligned}
    \zeta &= m\,\nu_{\max} + b \\
    \text{where} \quad
    m &= \slope \pm \slopeerr \ \mathrm{km\,s^{-1}\,\mu Hz^{-1}}, \\
    b &= \intercept \pm \intercepterr \ \mathrm{km\,s^{-1}}.
\end{aligned}
\label{macromodel}
\end{equation}

\noindent The two parameters have a scaled covariance of -0.011, which implies a correlation between slope and intercept of -0.94. Our relationship is restricted to use over the range of our data, \numaxxx ratio with solar value between 0.4 and 1.4, which includes many F/G/K dwarfs. 

We also fit a power law through this data, finding the scaling constant to be 2.89 $\pm$ 0.1 and the exponent to be -0.57 $\pm$ 0.07. The scatter around this fit is elevated to 0.19 \kms with $\chi^2_{red}$ of 1.0, and so we prefer the linear fit. For completeness, we also fit linear and power law $\zeta$-\numaxxx relationships through our full sample. In both cases, we find the same line as when modeling only the gold sample, within the overlapping 1$\sigma$ uncertainties, but higher RMS scatter and reduced $\chi^2$ values. Similarly, we repeated all models against the independent variable of \teff, see Table \ref{modelcompare}. Compared to their \numaxxx counterpart models, the gold sample \teff models have elevated scatter and RMS/reduced $\chi^2$. Meanwhile the full sample models perform marginally better than their \numaxxx counterparts, but all have elevated uncertainties on the model parameters due to the increased parameter uncertainty as a percentage of dynamic range of the independent variable, diminishing the practical usefulness of these models.

\begin{table}
\centering
\caption{Comparison of Models}
\label{modelcompare}
\setlength{\tabcolsep}{3.5pt}
\begin{tabular}{@{}lcccc@{}}
\toprule
Model & $a$ & $b$ & RMS & $\chi^2_{red}$ \\
\midrule
\multicolumn{5}{@{}c}{\numaxxx} \\
\textbf{Gold Linear}      & $2.30 \pm 0.12$ & $5.32 \pm 0.10$ & 0.09 & 0.26 \\
Full Linear      & $2.33 \pm 0.18$ & $5.36 \pm 0.15$ & 0.42 & 1.13 \\
Gold Power Law   & $2.89 \pm 0.10$ & $-0.58 \pm 0.07$ & 0.19 & 1.00 \\
Full Power Law   & $3.01 \pm 0.09$ & $-0.50 \pm 0.06$ & 0.51 & 2.07 \\
\midrule
\multicolumn{5}{@{}c}{\teff} \\
Gold Linear      & $0.0030 \pm 0.0006$ & $-13.9 \pm 3.4$ & 0.37 & 1.72 \\
Full Linear      & $0.0030 \pm 0.0003$ & $-13.6 \pm 1.7$ & 0.36 & 0.95 \\
Gold Power Law   & $3.64 \pm 0.14$ & $6.08 \pm 1.25$ & 0.34 & 1.41 \\
Full Power Law   & $3.57 \pm 0.06$ & $5.34 \pm 0.57$ & 0.40 & 0.96 \\
\bottomrule
\end{tabular}
\vspace{0.5em}
\begin{minipage}{\columnwidth}
\footnotesize
We adopt the bold model, Gold Linear against \numaxxx, which is reproduced in Equation~\ref{macromodel}.
All reported with units of \kms\ where applicable.
For Linear models, \textit{a} represents slope and \textit{b} represents intercept. For Power Law models, \textit{a} represents scaling factor and \textit{b} represents the exponent value.
\end{minipage}
\end{table}

The lower quality of the full sample models is due to the lower quality of the asteroseismic \vsin measurements in the wider sample. Our work relies on the veracity of the literature's \vsin measurement, which motivates fitting our relationship only through the higher quality asteroseismic data stars (as prescribed in \citet{Kamiaka2018}) making up the ``gold'' sample. Further highlighting this, across the full sample the uncertainties in our measured macroturbulence values correlate with the literature's \vsin uncertainty with a Pearson Correlation coefficient of 0.89. 

\section{Discussion}
\label{Discussion}

\subsection{\numaxxx vs. \teff}

In D14, the authors chose to fit their data as a power law function of \teff and \logg. Their work produced an RMS scatter of 0.37 {\kms} and they overall quote a 0.73 {\kms} uncertainty on a value determined from their relationship. Instead of applying \teff and \logg separately, we chose a combinatorial of the two parameters, \numaxxx, to serve as the sole input for our model. In this model, we find a tight linear relationship with an RMS scatter of just 0.09 km/s and a reduced $\chi^2$ of 0.26 that indicates our uncertainties are over-estimated by a factor of $\sim$2. This is in contrast to the model in \teff alone which has both an elevated RMS of 0.37 and reduced $\chi^2$ of 1.72. We find that \numaxxx has a greater predictive power of macroturbulence as it more fully captures the stellar structure.  

Physically, turbulence in stellar atmospheres is related to the convective motions in the outer layers of the star, which in turn depend on the pressure scale height and thus surface gravity. This is supported empirically by the strong observed correlation between surface gravity and temperature with the amplitudes of photometric variability from convective processes such as granulation \citep{Bastien2016,Kallinger2016}, which in turn are physically tied to atmospheric turbulence. Therefore, when two star’s \teff and \logg are nearly identical, they should have nearly identical macroturbulent broadening terms. Because both \teff and \logg are accounted for in the scaling relationship to \numaxxx, this parameter naturally traces macroturbulence more closely than \teff alone or the the D14 combination model.

\begin{figure}[t!]
\centering\includegraphics[width=0.5\textwidth]{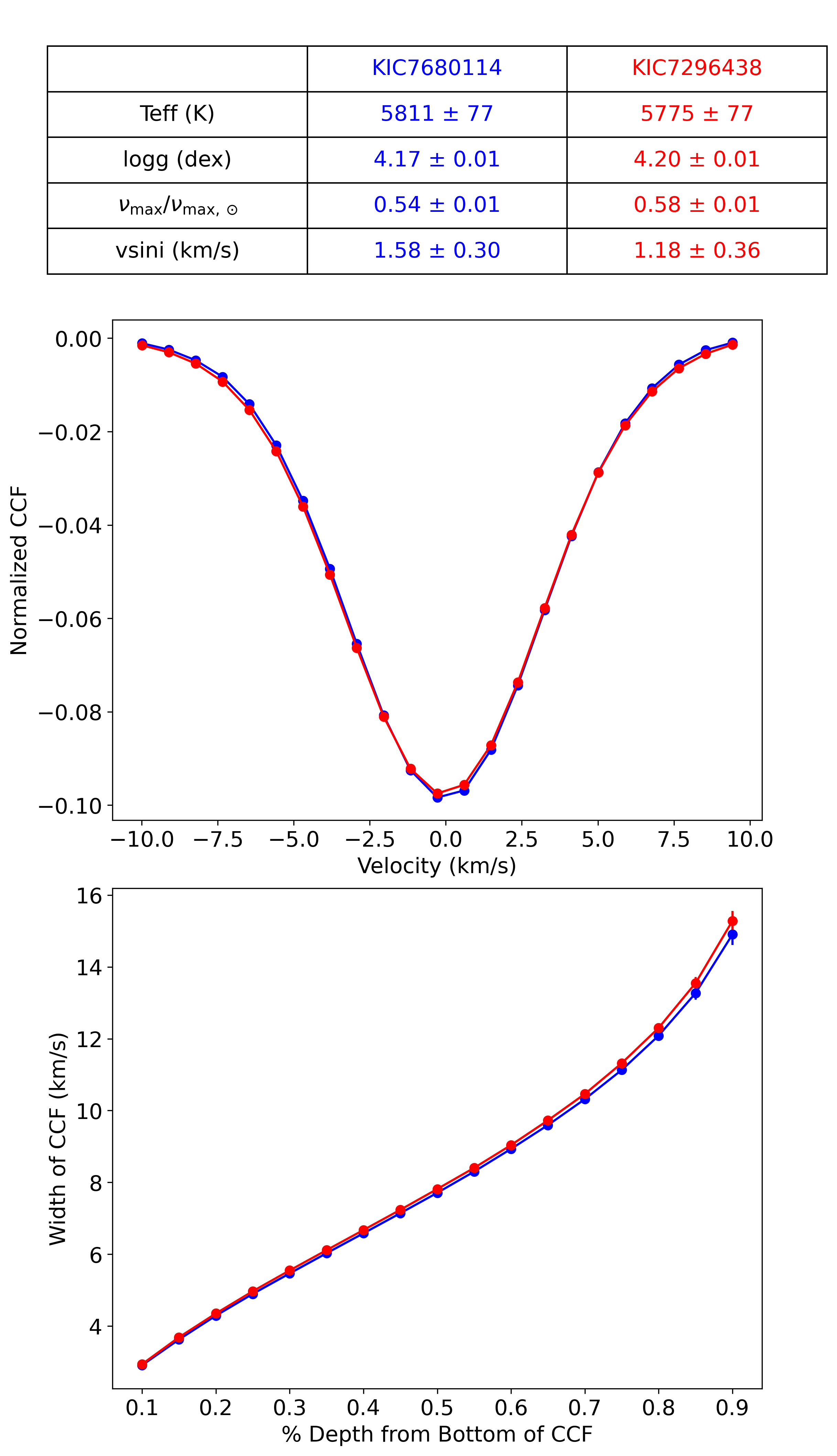}
\caption{\textbf{Top:} A table of stellar properties for two similar stars in our gold sub-sample. \textbf{Middle:} The KPF DRP's L2 weighted-sum CCFs of the Green chip for the two stars. \textbf{Bottom:} The widths of the two CCFs as a function of their depths. Note that uncertainties are included on both plots. Within macroturbulence and \vsin uncertainties, these stars have identical total broadenings, as expected due to their similar \numaxxx values.} 
\label{compare}
\end{figure}

Our width vs. depth curves further illustrate the relationship with \numaxxx. In our gold sub-sample, KIC7680114 and KIC7296438 have very close \numaxxx values and so they should have similar broadening from macroturbulence. Within the uncertainties of our $\zeta$-\numaxxx relationship, they should differ in macroturbulence broadening by only up to 0.14 {\kms}. They also have \vsin posteriors that overlap within 1$\sigma$. In the scenario which exists within both \vsin and $\zeta$ uncertainties where both values for both stars are exactly the same, we would expect the width vs. depth curves to overlap entirely. The actual curves do demonstrate a strong similarity, see Figure \ref{compare}. The average difference in CCF width values is 0.1 {\kms}, which, through comparison of synthetic data, would place the net difference between the sums of \vsin and $\zeta$ to be within 0.3 {\kms}, fully consistent with the parameter uncertainties.

\subsection{Future Work}
\label{Future}

Because we have modeled our relationship with a line, it is implied that at \numaxxx = $\sim$2, macroturbulence reaches zero and continues to fall into negative, nonphysical values. Stars in this regime are the M dwarfs; for example, the \numaxxx of GJ 687 is 2.15. This motivates us to restrict the validity of our relationship to the \numaxxx values spanning our data.

To motivate altering our relationship's functional form, we would require high quality asteroseismic \vsin measurements over a broader range of stellar \numaxxx values. This could be at either end of our current range. Including late K or early M dwarfs might indicate a minimum macroturbulence level that an exponential or step function captures better. Similarly, including evolved stars with lower \logg values for their \teff might also break the linear relationship and motivate a power law fit over the broader range of \numaxxx values. While the \textit{TESS} mission continues to extend baseline of photometry across the sky, including with very short cadence as low as 20 seconds, more stars may become amenable to asteroseismic mode-splitting measurements. That said, it is unlikely that \textit{TESS} will have the sensitively to add measurements for later dwarf stars. Regardless, we look forward to any additional slow-rotator benchmark stars that such a data set would produce.

The CCF is, in effect, the average line profile of the mask chosen to compute it. The line-by-line RV computation method \citep{Artigau2022} is a promising avenue of reducing RV jitter. It explores the use of different line lists like ones that single out lines of certain formation temperatures or optical depths \citep{AlMoulla2024, Palumbo2025} in an effort to compute less noisy RVs. Similar investigations into macroturbulence as a function of these kinds of line lists using our pipeline and methods may be fruitful. 

Additionally, one can imagine recreating this work with different or more complex broadening kernels, which may represent additional physics not modeled here. There are additional sources of broadening in stellar spectra that we have chosen to ignore. A more detailed treatment of thermal and/or pressure broadening and/or of magnetic fields and/or convective blueshift may produce even better results. 

The real insight of this work is that it sets up for a new way of leveraging the macroturbulence relationship. Given that the CCF is a low dimensional representation of a spectrum, it naturally extends itself to data driven techniques to infer information such as fundamental stellar parameters. It may be possible to train a data-driven model on CCFs to measure macroturbulence and/or slow \vsin accurately and precisely. Such a model would open up new avenues of study on the slow rotating, Sun-like stars that are of primary interest to the exoplanet sub-field. 

\section{Conclusion}
\label{Conclusion}

In this work, we have developed a synthetic spectral emulator for KPF that accounts for line broadening from macroturbulence, rotation, and the instrument itself, leveraging the instrument's stable profile. We further demonstrated that it reproduces results in measuring macroturbulence from traditional local thermodynamic equilibrium (LTE) spectral synthesis. 

We have measured macroturbulence through use of the CCF, a method independent to prior works that produces consistent, yet more precise results. Working with CCFs is common within the Extreme Precision Radial Velocity (EPRV) sub-field which is seeking to understand sub-{\ms} stellar activity and planetary RV signals. In this regime, macroturbulence represents a source of unwanted noise, so better understanding of how it manifests will aid in producing techniques to remove it before computing precision RVs. 
 
Lastly, we derived a linear relationship between a star's macroturbulence and \numaxxx, valid within the range of 0.4 to 1.4 \numaxxx as a ratio with the solar value. We further find that \numaxxx is strong predictor of macroturbulence. Such a relationship will be instrumental in measuring more precise \vsin values, in turn allowing for a more detailed study of many aspects of stellar and exoplanetary science.

\begin{table*}[ht]
\centering
\scriptsize
\setlength{\tabcolsep}{5pt}
\renewcommand{\arraystretch}{0.95}
\begin{tabular}{@{}lcccccccccccc@{}}
\toprule
Star & Gold$^a$ & KPF & $T_{\mathrm{eff}}^c$ & $\log g^c$ & [Fe/H]$^c$ & $u_1$ & $u_2$ & $\nu_s\sin i$$^c$ & $v\sin i^d$ & \numaxxx$^e$ & $\zeta$ This Work & $\zeta$ D14\\
 & & S/N$^b$ & K & dex & dex & & & $\mu$Hz & km\,s$^{-1}$ & $\mu$Hz & km\,s$^{-1}$ & km\,s$^{-1}$\\
\midrule
Sun & True & 586 & $5777$ & $4.44$ & $0.00$ & $0.50$ & $0.15$ & -- & $1.90 \pm 0.01^f$ & $1.00 \pm 0.00^g$ & $3.10 \pm 0.24$ & $3.20 \pm 0.51^h$ \\
KIC3544595 & False & 116 & $5669$ & $4.47$ & $-0.18$ & $0.50$ & $0.10$ & $0.40 \pm 0.04$ & $1.62 \pm 0.15$ & $1.08 \pm 0.01$ & $2.90 \pm 0.21$ &  \\
KIC3656476 & False & 150 & $5668$ & $4.23$ & $0.25$ & $0.59$ & $0.11$ & $0.21 \pm 0.02$ & $1.20 \pm 0.12$ & $0.62 \pm 0.01$ & $3.50 \pm 0.19$ & $3.56 \pm 0.49$ \\
KIC3735871 & False & 219 & $6107$ & $4.40$ & $-0.04$ & $0.51$ & $0.16$ & $0.69 \pm 0.05$ & $3.28 \pm 0.28$ & $0.88 \pm 0.01$ & $3.80 \pm 0.41$ &  \\
KIC4141376 & False & 146 & $6134$ & $4.41$ & $-0.24$ & $0.50$ & $0.17$ & $0.76 \pm 0.13$ & $3.45 \pm 0.61$ & $0.92 \pm 0.01$ & $4.50 \pm 0.96$ &  \\
KIC4914923 & True & 147 & $5805$ & $4.20$ & $0.08$ & $0.56$ & $0.13$ & $0.39 \pm 0.03$ & $2.32 \pm 0.18$ & $0.57 \pm 0.01$ & $4.10 \pm 0.22$ & $4.13 \pm 0.42$ \\
KIC5184732 & False & 56 & $5846$ & $4.25$ & $0.36$ & $0.57$ & $0.12$ & $0.55 \pm 0.02$ & $3.18 \pm 0.18$ & $0.65 \pm 0.01$ & $3.90 \pm 0.29$ & $3.67 \pm 0.64$ \\
KIC5950854 & False & 31 & $5853$ & $4.24$ & $-0.23$ & $0.53$ & $0.15$ & $0.29 \pm 0.38$ & $1.59 \pm 2.06$ & $0.63 \pm 0.01$ & $3.70 \pm 1.39$ &  \\
KIC6106415 & False & 38 & $6037$ & $4.30$ & $-0.04$ & $0.52$ & $0.15$ & $0.69 \pm 0.02$ & $3.69 \pm 0.15$ & $0.71 \pm 0.01$ & $3.90 \pm 0.30$ & $4.14 \pm 0.59$ \\
KIC6116048 & False & 133 & $6033$ & $4.25$ & $-0.23$ & $0.51$ & $0.15$ & $0.63 \pm 0.02$ & $3.27 \pm 0.18$ & $0.64 \pm 0.01$ & $4.10 \pm 0.33$ & $4.02 \pm 0.57$ \\
KIC6278762 & True & 117 & $5046$ & $4.56$ & $-0.37$ & $0.61$ & $0.12$ & $0.30 \pm 0.09$ & $0.98 \pm 0.28$ & $1.42 \pm 0.02$ & $2.10 \pm 0.48$ &  \\
KIC6521045 & True & 69 & $5825$ & $4.12$ & $0.02$ & $0.55$ & $0.13$ & $0.45 \pm 0.03$ & $2.95 \pm 0.17$ & $0.49 \pm 0.01$ & $4.20 \pm 0.24$ &  \\
KIC6933899 & False & 141 & $5832$ & $4.09$ & $-0.01$ & $0.55$ & $0.13$ & $0.36 \pm 0.02$ & $2.51 \pm 0.17$ & $0.44 \pm 0.01$ & $4.00 \pm 0.20$ & $4.19 \pm 0.63$ \\
KIC7296438 & True & 96 & $5775$ & $4.20$ & $0.19$ & $0.57$ & $0.12$ & $0.20 \pm 0.06$ & $1.18 \pm 0.36$ & $0.58 \pm 0.01$ & $4.10 \pm 0.21$ &  \\
KIC7680114 & True & 101 & $5811$ & $4.17$ & $0.05$ & $0.56$ & $0.13$ & $0.26 \pm 0.05$ & $1.58 \pm 0.30$ & $0.54 \pm 0.01$ & $3.90 \pm 0.18$ & $3.65 \pm 0.53$ \\
KIC7871531 & False & 26 & $5501$ & $4.48$ & $-0.26$ & $0.56$ & $0.14$ & $0.33 \pm 0.03$ & $1.24 \pm 0.12$ & $1.13 \pm 0.02$ & $2.70 \pm 0.29$ & $2.81 \pm 0.52$ \\
KIC7970740 & True & 244 & $5309$ & $4.54$ & $-0.49$ & $0.55$ & $0.15$ & $0.26 \pm 0.02$ & $0.86 \pm 0.08$ & $1.32 \pm 0.02$ & $2.20 \pm 0.19$ & $2.50 \pm 0.74$ \\
KIC8006161 & True & 215 & $5488$ & $4.49$ & $0.34$ & $0.62$ & $0.10$ & $0.34 \pm 0.02$ & $1.37 \pm 0.09$ & $1.17 \pm 0.02$ & $2.70 \pm 0.25$ & $2.22 \pm 0.58$ \\
KIC8150065 & False & 133 & $6173$ & $4.22$ & $-0.13$ & $0.51$ & $0.15$ & $0.54 \pm 0.12$ & $3.29 \pm 0.73$ & $0.59 \pm 0.01$ & $4.80 \pm 0.73$ &  \\
KIC9098294 & False & 99 & $5852$ & $4.31$ & $-0.18$ & $0.53$ & $0.15$ & $0.36 \pm 0.04$ & $1.81 \pm 0.19$ & $0.74 \pm 0.01$ & $4.20 \pm 0.19$ & $3.71 \pm 0.69$ \\
KIC9410862 & False & 28 & $6047$ & $4.30$ & $-0.31$ & $0.51$ & $0.16$ & $0.41 \pm 0.08$ & $2.08 \pm 0.43$ & $0.71 \pm 0.01$ & $3.70 \pm 0.48$ &  \\
KIC9955598 & False & 73 & $5457$ & $4.50$ & $0.05$ & $0.60$ & $0.12$ & $0.29 \pm 0.04$ & $1.13 \pm 0.15$ & $1.18 \pm 0.02$ & $2.50 \pm 0.18$ & $2.51 \pm 0.76$ \\
KIC10079226 & False & 156 & $5949$ & $4.37$ & $0.11$ & $0.54$ & $0.14$ & $0.65 \pm 0.07$ & $3.27 \pm 0.38$ & $0.84 \pm 0.01$ & $3.80 \pm 0.50$ &  \\
KIC10514430 & False & 33 & $5784$ & $4.06$ & $-0.11$ & $0.55$ & $0.13$ & $0.18 \pm 0.05$ & $1.26 \pm 0.72$ & $0.42 \pm 0.01$ & $4.20 \pm 0.34$ &  \\
KIC10516096 & False & 143 & $5964$ & $4.17$ & $-0.11$ & $0.53$ & $0.14$ & $0.48 \pm 0.03$ & $2.95 \pm 0.22$ & $0.53 \pm 0.01$ & $4.40 \pm 0.25$ &  \\
KIC10586004 & False & 29 & $5770$ & $4.07$ & $0.29$ & $0.58$ & $0.11$ & $0.48 \pm 0.17$ & $3.51 \pm 1.21$ & $0.43 \pm 0.01$ & $3.20 \pm 1.71$ &  \\
KIC11295426 & False & 29 & $5793$ & $4.28$ & $0.12$ & $0.56$ & $0.13$ & $0.22 \pm 0.03$ & $1.21 \pm 0.16$ & $0.70 \pm 0.01$ & $3.60 \pm 0.18$ &  \\
KIC11772920 & False & 85 & $5180$ & $4.50$ & $-0.09$ & $0.63$ & $0.11$ & $0.31 \pm 0.04$ & $1.13 \pm 0.14$ & $1.22 \pm 0.04$ & $2.00 \pm 0.26$ &  \\
KIC12069424 & True & 172 & $5825$ & $4.29$ & $0.10$ & $0.55$ & $0.14$ & $0.40 \pm 0.01$ & $2.12 \pm 0.07$ & $0.70 \pm 0.01$ & $3.70 \pm 0.14$ &  \\
KIC12069449 & False & 164 & $5750$ & $4.35$ & $0.05$ & $0.56$ & $0.14$ & $0.31 \pm 0.01$ & $1.46 \pm 0.07$ & $0.83 \pm 0.01$ & $3.60 \pm 0.14$ &  \\
\bottomrule
\multicolumn{13}{l}{\footnotesize All \vsin\ and $\zeta$ values are reported in units of {\kms}.} \\
\multicolumn{13}{l}{\footnotesize $a$ True if passes both cuts from \citet{Kamiaka2018}: HBR $>$ 3 and $\delta\nu/\Gamma > 0.5$, except for the Sun.} \\
\multicolumn{13}{l}{\footnotesize $b$ Measured via Median Absolute Deviation using Order 30.} \\
\multicolumn{13}{l}{\footnotesize $c$ Values reproduced from \citet{Hall2021} (except solar values).} \\
\multicolumn{13}{l}{\footnotesize $d$ Values and uncertainties computed using the mass, \logg, and $\nu_s\sin i$ from \citet{Hall2021}.} \\
\multicolumn{13}{l}{\footnotesize $e$ Computed via Equation~\ref{numaxEQ}.} \\
\multicolumn{13}{l}{\footnotesize $f$ Solar \vsin\ uncertainty was defined to be arbitrarily small in this work.} \\
\multicolumn{13}{l}{\footnotesize $g$ Solar \numaxxx\ is defined to be 1.0; for ODR we use the median \numaxxx\ uncertainty.} \\
\multicolumn{13}{l}{\footnotesize $h$ Solar $\zeta$ uncertainty is the average of the ESPaDOnS and Narval spectra used by D14.} \\
\end{tabular}
\caption{Macroturbulence measurements for all stars in our sample.}
\label{results_table}
\end{table*}

\section{Acknowledgments}
\label{Acknowledgements}

We recognize and acknowledge the cultural role and reverence that the summit of Maunakea has within the indigenous Hawaiian community. We are deeply grateful to have the opportunity to conduct observations from this mountain. 

J.L and E.P. are very grateful to the Heising-Simons Foundation for funding this work under grant number 2022-3833. K.M. acknowledges support from JSPS KAKENHI Grant Number 25K07387. D.H. acknowledges support from the National Aeronautics and Space Administration (80NSSC22K0781, 80NSSC25K7155) and NASA's Interdisciplinary Consortia for Astrobiology Research (NNH19ZDA001N-ICAR) under award number 19-ICAR19 2-0041.

The research was carried out, in part, at the Jet Propulsion Laboratory, California Institute of Technology, under a contract with the National Aeronautics and Space Administration (80NM0018D0004). 

\bibliography{main.bib}

\end{document}